\documentclass[prl,showpacs,preprint,floatfix]{revtex4}
\newcommand\FigOneScale{1.0}
\newcommand\FigTwoScale{1.0}
\newcommand\FigThreeScale{1.0}
\newcommand\FigFourScale{1.0}

\usepackage{graphics}

\begin{document}
\title{Asynchronous response of coupled pacemaker neurons}
\author{Ramana Dodla}
\author{Charles J. Wilson}
\affiliation{Department of Biology, University of Texas at San Antonio, San Antonio, TX 78249}

%
%
\begin{abstract}
We study a network model of two conductance-based pacemaker neurons of
differing natural frequency, coupled with either mutual excitation or
inhibition, and receiving shared random inhibitory synaptic input.  The
networks may phase-lock spike-to-spike for strong mutual coupling. But the
shared input can desynchronize the locked spike-pairs by selectively
eliminating the lagging spike or modulating its timing with respect to the
leading spike depending on their separation time window.  Such loss of
synchrony is also found in a large network of sparsely coupled heterogeneous
spiking neurons receiving shared input.
\end{abstract}

\maketitle

%
%

Phase locking or synchrony of oscillatory systems is a ubiquitous 
phenomenon in physics \cite{strogatz00from,hong07entrainment}, 
chemistry \cite{kuramoto84book}, and biology \cite{winfree80book,tsubo07synchronization}.
The theory of weakly coupled oscillators predicts that phase-locking can arise 
from mutual interactions among oscillators \cite{kuramoto87statistical,%
hansel95synchrony,mancilla07synchronization,lewis03dynamics,%
wang96gamma,tiesinga00synchronous,vanvreeswijk94when,jeong07synchrony}.
Phase locking is also facilitated by a common external 
input \cite{teramae04robustness,kawamura08collective} either with or 
without weak mutual interactions. But the theory of weakly coupled oscillators 
is strictly applicable only for oscillators with similar natural 
frequencies \cite{hoppenstead97book}. In some brain structures, there are
mutually-interconnected oscillatory neurons with more frequency heterogeneity
than allowed by the theory of weakly coupled oscillators \cite{bevan99mechanisms}.  
Oscillators with substantially different natural frequencies may still become 
phase-locked if coupled strongly enough \cite{matthews90phase}, but the 
influence of common inputs to these strongly coupled systems has not been fully 
explored \cite{white98synchronization,neltner00synchrony,binder01relationship,ly08synchronization}.
We are specifically motivated by studies of the subthalamic nucleus (STN) and globus
pallidus (GP) which contain interconnecting oscillatory neurons that receive
common inhibitory inputs.  These neurons are normally uncorrelated but become
strongly correlated in Parkinson's disease \cite{raz00firing}.  In this Letter
we construct oscillating two-neuron network models with different natural
frequencies but that are phase-locked due to strong mutual interconnections.
We show that a common external inhibitory input can reduce or abolish
phase-locking and decorrelate their firing. The amount of decorrelation depends
on the phase difference between the spike times in the phase-locked state.

Our results are demonstrated with coupled two-neuron networks that use three
different models of varying complexity: a detailed neuron model that reflects
realistic properties of the STN neurons including rebound currents, a reduced
three-variable model (N+K) derived from the first model, and the generic Hodgkin
Huxley (HH) neuron model, which was also used as a model of the GP neuron. The
results are also extended to a larger network of 10 sparsely coupled neurons.
We used a shared external inhibitory input, because both the GP and STN receive
strong inhibitory inputs. Interconnections among neurons within the STN
network model are excitatory, whereas the GP cells are coupled by inhibitory
connections.  Unlike integrate-and-fire type neuron models that may fail to
synchronize for strong coupling \cite{golomb00number} 
or for common input \cite{teramae04robustness}, our neuron pairs can
display robust 1:1 phase-locking of spike times in the absence of external
input. This kind of 1:1 phase-locking is sometimes also called linear
synchrony \cite{schiff96detecting,pyragas96weak} in studies of coupled chaotic 
oscillators. Our study highlights the importance
of the timing of the arriving inputs in decorrelating the phase-locked state.
However, we confine our study to neuron models that are periodic and are not chaotic.

We consider two coupled pacemaking model neurons at dissimilar frequencies
($\omega_{1,2}$) and affected by identical synaptic noise.
The current balance equations (with unit membrane capacitance) are
\begin{eqnarray}
\dot{V}_1=I_1 - I^1_{g} - g S_2 \times (V_1-E) - I_{\mathrm{inh}}(V_1,t), \nonumber \\
\dot{V}_2=I_2 - I^2_{g} - g S_1 \times (V_2-E) - I_{\mathrm{inh}}(V_2,t), \nonumber
\end{eqnarray}
where $V_{1,2}$ are the membrane voltages. 
For mutual excitation, $E \equiv E_{\mathrm{exc}} =  10$ mV and $g \equiv g_{\mathrm{exc}}$,
and for mutual inhibition $E \equiv E_{\mathrm{inh}} = -85$ mV and $g \equiv g_{\mathrm{inh}}$.
$I_{j}$ is a steady applied current. $j=1,2$ is the index of the neuron.
All the three models were used with mutual
excitation, and the HH model was also used with mutual inhibition.
For the realistic STN model, $I^j_g\equiv I_g(V_j,n_j,h_j,r_j,[\mathrm{Ca}]_j)$, is a function of
$V_j$, potassium activation ($n_j$), sodium inactivation ($h_j$), low-threshold
calcium inactivation ($r_j$), and calcium concentration ($[\mathrm{Ca}]_j$) variables.
These two equations along with those for $n_j$, $h_j$, $r_j$, and $[\mathrm{Ca}]_j$,
together describe the 
coupled system \footnote{See
\cite{terman02activity} for the original formulation.  $dx_j/dt=(x_\infty(V_j)
-x_j)/\tau_x(V_j)$, where $x \equiv n, h,$ and $r$; $\frac{d[\mathrm{Ca}]_j}{dt} =
3.75\times10^{-5} (-I_{\mathrm{Ca}}(V_j) - I_T(V_j,r_j) - 22.5 [\mathrm{Ca}]_j),$
where
$I_g(V_j,n_j,h_j,r_j,[\mathrm{Ca}]_j) =  
               I_{\mathrm{L}}(V_j) +
               I_{\mathrm{Na}}(V_j,h_j) +
               I_K(V_j,n_j) +
               I_{\mathrm{Ca}}(V_j)  +
               I_T(V_j,r_j) +
               I_{\mathrm{AHP}}(V_j,[\mathrm{Ca}]_j)$,
and
$I_{\mathrm{Na}}(V,h)    = G_{\mathrm{Na}}  {m}_\infty^3(V) h            \times (V-E_{\mathrm{Na}})$, 
$I_{\mathrm{K}}(V,n)     = G_{\mathrm{K}}     n^4                          \times (V-E_\mathrm{K})$, 
$I_{\mathrm{L}}(V)       = G_{\mathrm{L}}                                  \times (V-E_\mathrm{L})$, 
$I_{\mathrm{Ca}}(V)      = G_{\mathrm{Ca}}  s_\infty^2(V)                \times (V-E_{\mathrm{Ca}})$, 
$I_{\mathrm{T}}(V,r)     = G_{\mathrm{T}}     a_\infty^3(V) b_\infty^2(r)  \times (V-E_{\mathrm{Ca}})$,
$I_{\mathrm{AHP}}(V,[\mathrm{Ca}]) = G_{\mathrm{AHP}} {[\mathrm{Ca}]/([\mathrm{Ca}]+15)}              \times (V-E_K).$
$m_\infty(V) = 1/[1+e^{-(V+30)/15}]$,
$h_\infty(V) = 1/[1+e^{-(V+39)/3.1}]$.
$n_\infty(V) = 1/[1+e^{-(V+32)/8}]$,
$s_\infty(V) = 1/[1+e^{-(V+39)/8.0}]$.
$r_\infty (V)= 1/[1+e^{(V+67)/2}]$,
$a_\infty(V) = 1/[1+e^{-(V+63)/7.8}]$, and
$b_\infty(r) = 1/[1+e^{-10(r-0.4)}] - 1/[1+e^{4}]$.
$\tau_n(V)   = \frac{4}{3} [1 + \frac{100}{1+e^{(V+80)/26}}]$,
$\tau_h(V)   = \frac{4}{3} [1 + \frac{500}{1+e^{(V+57)/3}}]$,
$\tau_r(V)   = 5 [40 + \frac{17.5}{1+e^{(V-68)/2.2}}]$.
$E_{\mathrm{L}} = -60$ mV,
$E_{\mathrm{K}} = -80$ mV,
$E_{\mathrm{Na}} = 55$ mV, and
$E_{\mathrm{Ca}} = 140$ mV.
$G_\mathrm{L} = 2.25$ nS/$\mu$m$^2$,
$G_\mathrm{K} = 45$ nS/$\mu$m$^2$,
$G_{\mathrm{Ca}} = 0.5$ nS/$\mu$m$^2$,
$G_{\mathrm{Na}} = 37.5$ nS/$\mu$m$^2$,
$G_\mathrm{T}    = 0.5$ nS/$\mu$m$^2$, and
$G_{\mathrm{AHP}}= 9.0$ nS/$\mu$m$^2$.}.
For the HH model, $I_g^j \equiv I_j(V_j,m_j,h_j,n_j)$, and
the gating variables are given by standard equations \cite{koch99book} 
at $T=0^o$C, $Q_{10}=3^{(T-6.3)/10}$.
A transformation $V\rightarrow -V-60$ was used, 
and $E_\mathrm{L}$ was depolarized from $-49.387$ mV to $-17$ mV 
to trigger pacemaking. The N+K model is derived from the STN by setting 
rebound ($G_\mathrm{T}$) and all other calcium
dependent maximal conductances ($G_{\mathrm{Ca}}, G_{\mathrm{AHP}}$) to zero, and thus
$I_g^j\equiv I_j(V_j,n_j,h_j)$.
The conductance state variable, $S_{1,2}\equiv S(V_{1,2},t)$ is noninstantaneous
and depends on the time history of $V_{1,2}$ via a differential equation \cite{wang92alternating}, 
and can range from $0$, when the membrane is near rest, to $1$ during an action potential %
\footnote{%
$S(V_j,t)$ evolves according to
$dS(V_j,t)/dt=4/(1+e^{-(V_j-\theta_{\mathrm{syn}})/k_{\mathrm{syn}}})
[1-S(V_j,t)]-\beta S(V_j,t)$, where $\theta_{\mathrm{syn}}=-20$ mV,
$k_{\mathrm{syn}}=2$ mV. When unspecified, $\beta=2$ ms$^{-1}$ in all networks.}. 
This incorporates a mutual coupling time constant, $1/\beta$.
$I_{\mathrm{inh}}(V_j,t)$ is an inhibitory input current generated from 
alpha function conductances timed at random but 
identical Poisson process arrival times ($t_i, i=1,\ldots$) with rate 
$\lambda_{\mathrm{inh}}$: $I_{\mathrm{inh}}(V_j,t) = G_{\mathrm{inh}} \sum_{t_i}
\frac{t-t_i}{\tau_{\mathrm{inh}}} e^{1-\frac{t-t_i}{\tau_{\mathrm{inh}}}}
H(t-t_i) \times (V_j-E_{\mathrm{inh}}).$ 
$G_{\mathrm{inh}}=0-1$ nS/$\mu$m$^2$ for 
STN and N+K, and $0-1$ mS/cm$^2$ for the HH network, and
$\tau_{\mathrm{inh}}=1-10$ ms.  
A fourth order Runge-Kutta algorithm with $1$ $\mu$s time
step was used to integrate up to $1000$ s. 
The spike time sequences ($s_1$ and $s_2$) of our networks, when they were not driven, 
are 1:1 phase-locked with spike separation $t_w$
and thus fall in the $(s_1,s_2)$ phase space on the diagonal line $s_1=s_2$.
The deviations from this linear synchrony (also see \cite{schiff96detecting} and references there in)
due to shared input are large and random. An asynchrony parameter that measures the
fraction of such deviating spike times is defined:
$A = 1 - N_{\mathrm{sync}}/N_{\mathrm{max}}$, where
$N_{\mathrm{sync}}$ is the number of spike time pairs (one spike from each neuron) in a time
window $t_w$
and $N_{\mathrm{max}}=\textrm{max}\{N_1,N_2\}$, where $N_{1,2}$ are spike counts of the two neurons.
$A=0$ for 1:1 phase-locked state.
For phase-locked states that have large $t_w$ [in Fig.~\ref{fig3}(b)], 
shorter separations also constitute deviations from the coherent state.
Thus an asynchrony measure called inter spike interval (ISI) distance \cite{kreuz07measuring} 
defined as $\frac{1}{T}\int_0^T\frac{\mid s_1-s_2\mid}{\mathrm{max}\{s_1,s_2\}} dt$ is used.
The ISI-distance quantifies the jitter around the diagonal line, $s_1=s_2$. 
However, both these measures give qualitatively similar results.

\begin{figure}
\centerline{\scalebox{\FigOneScale}{\includegraphics{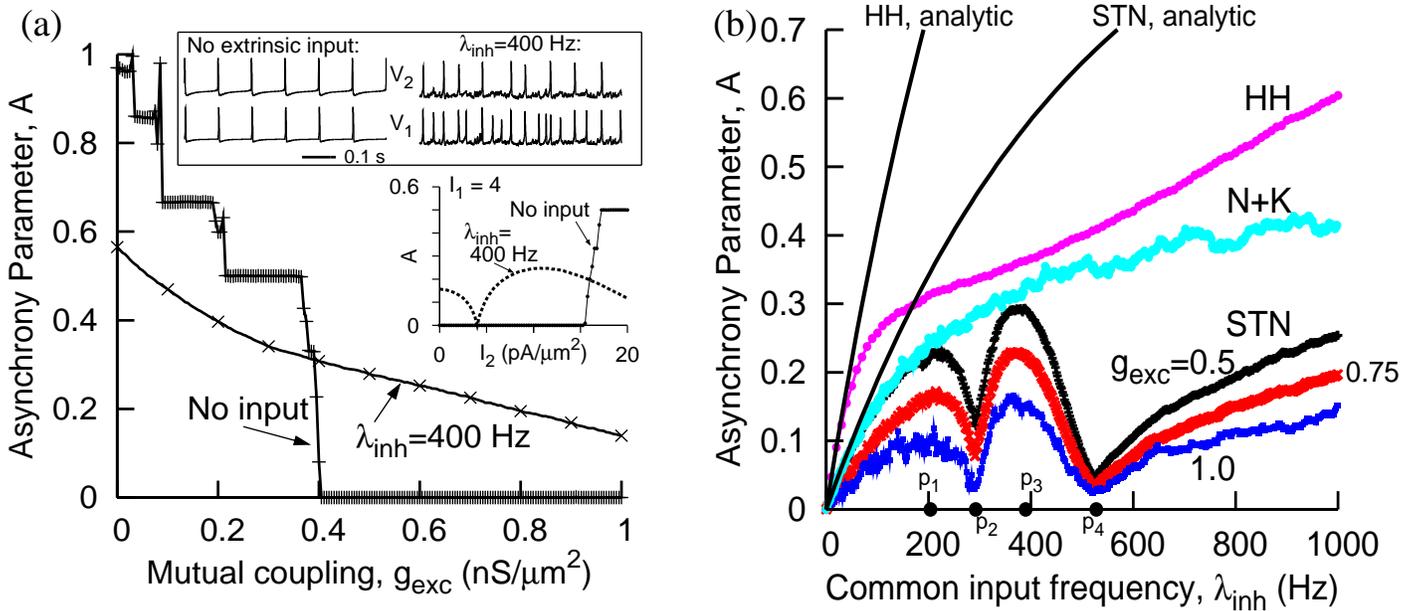}}}
\caption{Emergence of asynchrony for shared inhibition.
(a) Coupled STN network.
$\omega_1=2.7$ Hz ($I_1=0$) and $\omega_2=9.3$ Hz 
($I_2=8$ pA/$\mu$m$^2$). $G_\mathrm{inh}=1$ nS/$\mu$m$^2$. 
$g_{\mathrm{exc}}^* = 0.4$ nS/$\mu$m$^2$.
The top inset shows brief time courses at $g_{\mathrm{exc}}=0.5$ nS/$\mu$m$^2$.
For the same conductance parameters, the effect of frequency heterogeneity 
is shown in the bottom inset.
(b) Frequency dependence of $A$
for the three networks: STN ($G_{\mathrm{inh}}=1$ nS/$\mu$m$^2$, 
and three different values of $g_{\mathrm{exc}}$),
N+K ($I_1=0$, $I_2=8$ pA/$\mu$m$^2$, $g_{\mathrm{exc}}=0.62$ nS/$\mu$m$^2$, $G_{\mathrm{inh}}=1$ nS/$\mu$m$^2$), 
and HH ($I_1=0$, $I_2=8$ $\mu$A/cm$^2$, $g_{\mathrm{exc}}=0.2$ mS/cm$^2$, $G_{\mathrm{inh}}=1$ mS/cm$^2$).
$p_1$, \ldots, $p_4$ mark input frequencies at which $A$ changed its character 
in the STN network due to T-current activation, and are described in the text.
($\tau_{\mathrm{inh}}=1$ ms). 
\label{fig1}}
\end{figure}

Figure~\ref{fig1}(a) shows the results of two mutually excitatory STN neurons that
have very different intrinsic spike frequencies. In the absence of external input,
mutual coupling that exceeds a critical level ($g_{\mathrm{exc}}^*$) phase-locks the
spike times of both the neurons in 1:1 ratio. This causes $A$ to become $0$
as the frequencies of both the cells become identical. An example of the
voltage time courses in this state is shown in Fig.~\ref{fig1}(a), top inset, left.
When the external inhibitory input was turned on [Fig.~\ref{fig1}(a),
top inset, right], the spike times of both the cells became arrhythmic and
the firing rates became different. The common input countered the effect of mutual
coupling by decreasing the number of the 1:1 phase-locked spike pairs.
The enhanced spike count resulted due to activation of low-threshold
rebound calcium current (i.e. the T-current). But the loss of phase-locking 
is not dependent on the T-current activation (see results using the N+K network below).
The result of the loss of the 1:1 phase-locking between the spike
times is also seen as a finite non-zero value of $A$ in the earlier
phase-locked regime ($g>g_{\mathrm{exc}}^*$). 
For this network, frequency heterogeneity is essential to elicit finite $A$.
This is demonstrated in Fig.~\ref{fig1}(a), bottom inset. 
In the absence of input, phase-locking was
achieved for a wide range of frequency disparity.
With inhibitory input, $A$ became
non-zero in this region except when the frequencies were identical ($I_1=I_2$).
The dependence of $A$ on the input rate is shown in Fig.~\ref{fig1}(b) for all the
two-neuron coupled model networks with mutual excitation.  In the absence of
input (i.e. at $\lambda_{\mathrm{inh}}=0$), the networks were 1:1 phase-locked
(i.e. $A=0$). At an input rate of $1$ kHz, about 40\% of the locked spikes in
the N+K network, and more than 50\% of the locked spikes in the HH network became
unpaired. But the STN network model exhibited frequency-dependent activation 
of its rebound current that led to a modulation of $A$. 
The T-current is most actively recruited in the slower neuron 
between $p_1$ and $p_3$, and in the faster neuron between $p_2$ and $p_4$. 
Thus, the firing rate as a function of $\lambda_{\mathrm{inh}}$ increased in both the neurons
between $p_2$ and $p_3$ leading to an increasing $A$.
Active recruitment in only one of the cells led to enhancement of phase-locking or reduction in $A$.

\begin{figure}
\centerline{\scalebox{\FigTwoScale}{\includegraphics{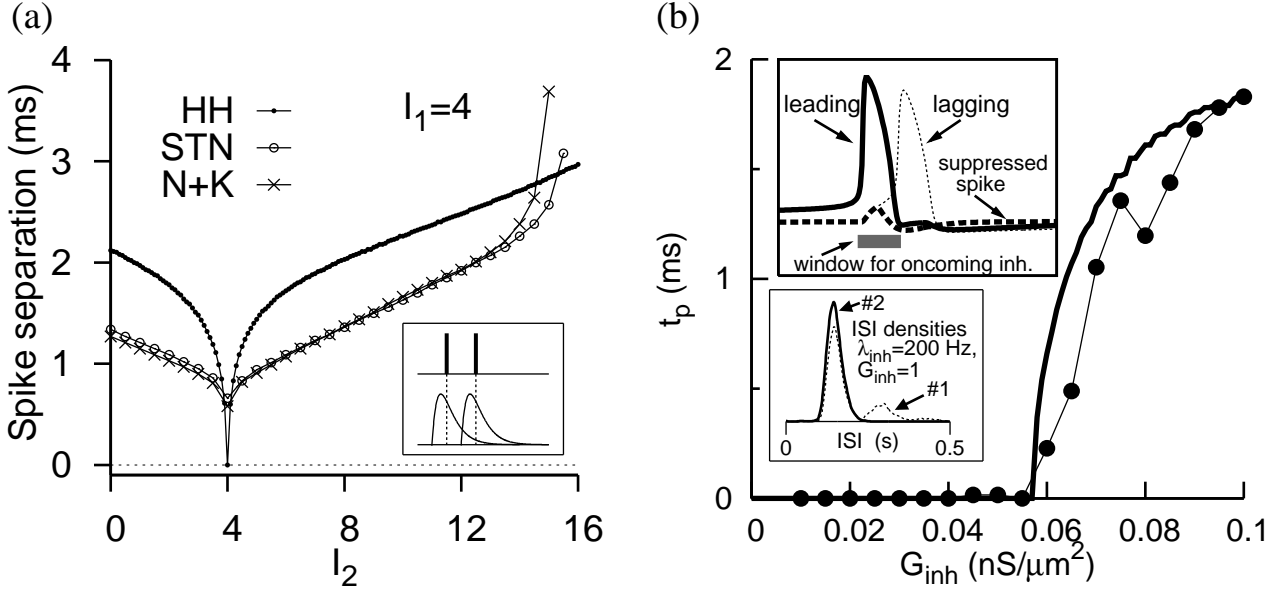}}}
\caption{(a) Time window of spike separation in 1:1 phase-locked state
in the absence of extrinsic input for the coupled STN, N+K and HH excitatory neuronal pairs 
as a function of frequency heterogeneity ($I_1$ is fixed at $4$ pA/$\mu$m$^2$ for STN
and N+K, and at $4$ $\mu$A/cm$^2$ for the HH.) The inset shows scenarios of input
events with respect to spike pair separation discussed in the text.
(b) Excitatory STN network. Predicted and computed time windows (see text).
The top inset shows the mechanism by which oncoming inhibition suppressed a lagging spike.
The bottom inset shows the resultant interspike interval histogram densities.
($\tau_{\mathrm{inh}}=1$ ms, $g_{\mathrm{exc}}=0.5$ nS/$\mu$m$^2$.)
\label{fig2}}
\end{figure}

The loss of 1:1 phase-locking observed in the N+K, the HH, and the
STN network (for $\lambda_{\mathrm{inh}}<p_1$, hence no significant rebound
spikes) emerges by selective elimination of one of the spikes in otherwise
phase-locked spike pairs [see ISI histogram in
Fig.~\ref{fig2}(b) bottom inset].  
Shared input cannot desynchronize the locked state if the spike pair separation $t_w=0$. 
For small $t_w$, a brief synaptic input is needed for disruption of the locked state.
$t_w$ is determined by the dynamics of the synaptic coupling conductance ($S(V_j,t)$) 
and the coupling strength $g_{\mathrm{exc}}$. 
It increases with frequency heterogeneity [Fig.~\ref{fig2}(a)].
For the STN network, empirically, $t_w \sim a (g_{\mathrm{exc}}-g_{\mathrm{exc}}^*)^{-b}$, 
where $a=0.78$, $b=0.44$
away from the criticality ($g_{\mathrm{exc}}^*$) , and $a=1.3$, $b=0.2$ near
criticality.  An inhibitory stimulus event preceding a locked pair
[Fig.~\ref{fig2}(a) inset] may reduce the spike separation, but one or more
events that are positioned appropriately in a time window $t_p$ (smaller or
bigger than $t_w$) with respect to, but preceding the onset of the excitatory postsynaptic conductance, can
eliminate the stimulated spike in the lagging neuron.  $A$ is a measure of the
probability of such a favorable window receiving one or more inhibitory stimuli
($1-P_0(\lambda_{\mathrm{inh}},t_p)$, where $P_n(\lambda,t)$ is the Poisson probability
density) that successfully eliminate the following spike.  Assuming a
successful elimination of the secondary spike by one or more arriving inputs in
a window, and assuming $t_p\approx t_w$, we write $A = Q (1-e^{-\lambda_{\mathrm{inh}}
t_w})$, where $Q$ is a constant that depends on the phase-locked frequency.  
$dA/d\lambda_{\mathrm{inh}} = Q t_w
e^{-\lambda_{\mathrm{inh}} t_w}$.  $Q$ was estimated 
numerically from $dA/d\lambda_{\mathrm{inh}}$. 
At $G_{\mathrm{inh}}=0.1$ nS/$\mu$m$^2$ and $t_w=1.8$ ms, $Q=0.034$.
By using $Q$, and the numerically evaluated $dA/d\lambda_{\mathrm{inh}}$, we solved the above
transcendental equation for $t_w$ at several $G_{\mathrm{inh}}$ values, and plotted
it as $t_p$ in Fig.~\ref{fig2}(b), in dots . 
A similar window for suppression of the lagging spike is computed deterministically
by using an inhibitory conductance input preceding the phase-locked spike pair, and
is shown in Fig.~\ref{fig2}(b) (thick solid line) for comparison.
Alternatively, $A$ may be approximated.
Such an approximation even at large $G_{\mathrm{inh}}$ values well predicts the growth
of $A$ [Fig.~\ref{fig1}(b), STN, analytic] at small $\lambda_{\mathrm{inh}}$:
$A=1.02~(1-e^{-0.002 \lambda_{\mathrm{inh}}})$. Using a similar approximation for the HH
network [Fig.~\ref{fig1}(b), HH, analytic] yields $A=1.8~(1-e^{-0.0026 \lambda_{\mathrm{inh}}})$.
For basal ganglia networks, values of $\tau_{\mathrm{inh}}$ are near $10$ ms. 
Increasing $\tau_{\mathrm{inh}}$ generally decreased the firing rates in the models 
because the input turned increasingly deterministic, and
the asynchrony parameter showed a moderate increment [Fig.~\ref{fig3}(a)]. 
But the STN network showed up to $80\%$ loss of phase-locked spike pairs
when $G_{\mathrm{inh}}$ is reduced appropriately at long $\tau_\mathrm{inh}$
such that the cells could maintain high firing rates due to activated T-current.
The modulation in $A$ is similar in nature to that seen in Fig.~\ref{fig1}(b). 

\begin{figure}
\centerline{\scalebox{\FigThreeScale}{\includegraphics{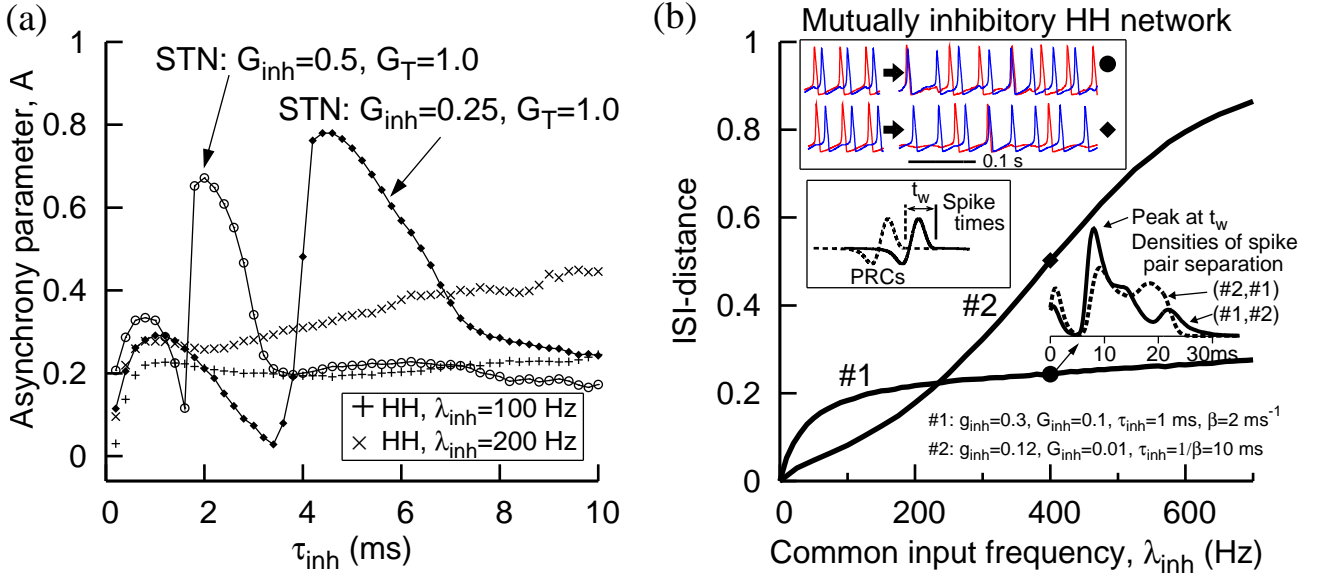}}}
\caption{(a) Dependence of $A$ on $\tau_{\mathrm{inh}}$.
For STN network $\lambda_{\mathrm{inh}}=100$ Hz, $g_{\mathrm{exc}}=0.5$ nS/$\mu$m$^2$.
For HH network $G_{\mathrm{inh}}=0.5$ mS/cm$^2$ and $g_{\mathrm{exc}}=0.1$ mS/cm$^2$.
$I_1$ and $I_2$ are as in Fig.~\ref{fig1}(b). $1/\beta=3$ ms for both the networks.
(b) Mutually inhibitory network of two coupled HH neurons
($I_1=0$, $I_2=8$ $\mu$A/cm$^2$) displaying spike time asynchrony 
for inhibitory input. 
At $\lambda_{\mathrm{inh}}=0$, $t_w=7.8$ for \#1 and $9.6$ ms for \#2.  
Voltage traces transitioning from phase-locked ($\lambda_{\mathrm{inh}}=0$) to
asynchronous state for finite input rate are illustrated in the top inset.
The phase response curves corresponding to the two phase-locked neurons 
are illustrated in the left inset, bottom. The right inset shows
the densities of spike pair separation in the asynchronous state.
\label{fig3}}
\end{figure}

\begin{figure}
\centerline{\scalebox{\FigFourScale}{\includegraphics{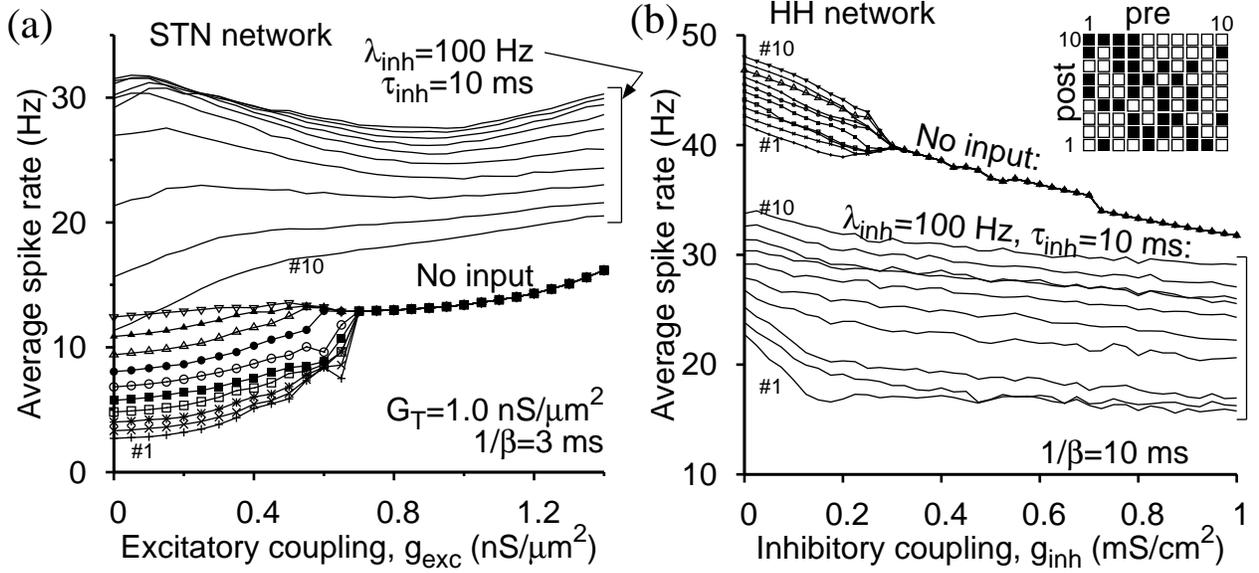}}}
\caption{Emergence of asynchrony from coherent states in networks of $N=10$
excitatory all-to-all coupled STN (a) and sparsely coupled inhibitory HH (b)
neurons measured by average spike rate as a function of mutual coupling without
and with shared inhibition.  (a) $I_j=(j-1) 10/9$ pA/$\mu$m$^2$,
$j=1,\ldots, N$, and $G_{\mathrm{inh}}= 0.5$ nS/$\mu$m$^2$.
(b) Inset shows $M_{ij}$ (with $n=4$); a filled square represents a 
synaptic connection. $I_j=5+(j-1) 8/9$$\mu$A/cm$^2$,
$j=1,\ldots, N$, $G_{\mathrm{inh}}= 0.1$ mS/cm$^2$.
\label{fig4}}
\end{figure}

We now briefly discuss the behavior of mutually inhibitory HH neurons [Fig.~\ref{fig3}(b)]
in response to common external Poisson inhibition.
Phase-locking can now occur in such networks with $t_w$ near the anti-phase state.
$t_w$ is modified by the frequency heterogeneity.
This phase difference between the spikes is either decreased or
increased by the occurrence of oncoming inhibitory synaptic conductance events
before or in between the phase-locked spike pair. 
This increased the asynchrony as measured by the ISI-distance (defined earlier)
with increasing $\lambda_{\mathrm{inh}}$ at fast as well as slow synaptic time constants,
$1/\beta$.
Finally, we illustrate the manifestation of these two forms of asynchrony in a
larger network ($N=10$) of excitatory [STN, all-to-all coupled; $n=9$;
Fig.~\ref{fig4}(a)] and inhibitory [HH, sparse coupled; $n=4$;
Fig.~\ref{fig4}(b)] neurons. The current balance equations are $\dot{V}_j=I_j -
I^j_{g} - g \frac{1}{n} \sum_{k=1}^{N} M_{jk} S_k \times (V_j-E) -
I_{\mathrm{\mathrm{inh}}}(V_j,t)$, $j=1,\ldots, 10$.  $M_{jj}=0$. $M_{jk} = 1,$ if the
presynaptic neuron $k$ is connected to the postsynaptic neuron $j$, and $0$
otherwise. In both the networks, in the absence of the input, the spike times were asynchronous
for weak coupling (see \cite{chik04clustering} for T-current induced clustering),
but strong mutual coupling above a critical level
synchronized the average firing rates of all the neurons to a common
frequency. When the shared inhibition is turned on, the STN network
neurons fired faster due to T-current activation, and the HH network neurons fired
slower. But the average firing rates exhibited asynchrony by diverging 
from the common locked frequency.

In contrast to the earlier studies on weakly coupled oscillator theory,
we showed that in strongly coupled heterogeneous networks, shared 
inhibition can lead to asynchrony of spike times. Our simulations
also revealed similar role for shared excitation.
Our results lead to the possibility that modulating
the synaptic input strength or input frequency might lead to switches in the
synchrony-asynchrony (or coherence-incoherence) transitions in 
coupled oscillatory networks that receive coherent input. 
It would be of interest to study the emergence of
asynchrony due to shared inputs when the phase-locked state is affected by
background noise \cite{yu07frequency} or if the oscillators themselves behave 
chaotically \cite{rosenblum97phase}. In Parkinsonian network models of deep brain stimulation,
earlier studies \cite{rubin04high,feng07optimal} relied on the network 
connectivity between different brain nuclei to produce complex asynchronous
patterns. Our results suggest of an alternative mechanism for generating such asynchrony 
within each brain nucleus. 

We thank the Computational Biology Initiative
(UTHSCSA/UTSA) and Texas Advanced Computing Center (TACC), The University of
Texas at Austin for providing high performance computing resources. 
The work was supported by NIH/NINDS NS47085.


\end{document}